\documentclass[aps,pre,amsmath,amssymb,groupedaddress]{revtex4}
\usepackage{graphicx}

\newcommand{\re}[1]{(\ref{#1})}
\newcommand{\ci}[1]{\cite{#1}}
\renewcommand{\baselinestretch}{1.2}
\begin{document}%\large
\title{Stationary Nonlinear Schr\"odinger Equation on Simplest Graphs:\\ Boundary conditions and exact solutions}
\author{Z.A. Sobirov, K.K. Sabirov and D.U. Matrasulov}\thanks{E-mail: d.matrasulov@polito.uz}
\affiliation{Turin Polytechnic University in Tashkent, 17. Niyazov
Str., 100095, Tashkent, Uzbekistan}

\begin{abstract}

We treat the stationary (cubic) nonlinear Schr\"odinger equation
(NSLE) on simplest graphs. Formulation of
the problem and exact analytical solutions of NLSE are presented
for star graphs consisting of three bonds. It is shown that the
method can be extended for the case of arbitrary number of bonds
of star graphs and for other simplest topologies such as tree and loop
graphs. The case of repulsive and attractive nonlinearities are
treated separately.
\end{abstract}

\maketitle PACS: 05.45.Yv, 42.65.Tg, 42.65.Wi, 03.75.-b, 05.45.-a, 05.60.Gg.
\section{Introduction.}

The nonlinear Schr\"odinger equation has attracted much attention
since from its pioneering studies in early seventies of the last
century \cite{Zakh1}-\cite{Zakh3}. Such attention was caused by
the possibility for obtaining soliton solution of NLSE and its
numerous applications in different branches of physics. The early
applications of NLSE and other nonlinear PDEs having soliton
solutions were mainly focussed in optics, acoustics, particle
physics, hydrodynamics and biophysics.  However, special attention
NLSE and its soliton solutions have attracted because of the recent
progress made in the physics and Bose-Einstein condensates(BEC).
Namely, due to the fact that the dynamics of BEC is governed by
Gross-Pitaevski equation which is NLSE with cubic nonlinearity,
finding the soliton solution of NLSE with different confining
potentials and boundary conditions is of importance for this area of physics.

Many aspects of soliton solution of NLSE have been treated during
the past decade in the context of fiber optics, photonic crystals,
acoustics and BEC (see books \cite{Kivshar1}-\cite{Thierry} and
references therein). Both, stationary and time-dependent NLSE
were extensively studied for different trapping potentials in the context of BEC.
In particular, the stationary NLSE was studied for
box boundary conditions \cite{Car1,Car2} and the square well
potential \cite{Well1}-\cite{Infeld}.

In this paper we  treat the stationary NLSE on graphs.  Graphs are
the systems consisting of {\underline bonds} which are connected
at the {\underline vertices} \cite{Harary}. The bonds are
connected according to a rule that is called topology of a graph.
Topology of a graph is given in terms of so-called adjacency
matrix (or connectivity matrix) which can be written as
\cite{Uzy1,Uzy2}:
\begin{equation}
C_{ij}=C_{ji}=\left\{\begin{array}{ccl}&1 & \mbox{ if }\; i\;
\mbox{ and }\; j\; \mbox{ are connected, }\\
& 0 & \mbox{ otherwise, }\end{array}\right. \qquad
i,j=1,2,...,V.\nonumber
\end{equation}

The linear Schr\"odinger equation on graphs has been topic of
extensive research recently (e.g., see review
\ci{Uzy1}-\cite{Uzy3} and references therein). In this case
the eigenvalue problem is given in terms of the boundary conditions
providing continuity and current conservation
\ci{Uzy1}-\cite{Exner2}.

Despite the progress made in the study of linear Schr\"odinger
equation on graphs, corresponding nonlinear problem, i.e., NLSE on
graphs is still remaining as less-studied problem. This is mainly
caused the difficulties that appear in the case of NLSE
on graphs, especially, for the time-dependent problem. In
particular, the problem becomes rather nontrivial and it is not so
easy to derive conservation laws \ci{Zarif}. It should be noted that during the last couple of years there were some
attempts to treat time-dependent \cite{Zarif, Adami} and
the stationary \cite{Cascaval,Uzy4} NLSE on graphs. Soliton solutions
and connection formulae are derived for simple graphs in the
Ref.\cite{Zarif}. The problem of fast solitons on star graphs is
treated in the Ref.\cite{Adami}. In particular, the estimates for
the transmission and  reflection coefficients are obtained in the
limit of very high velocities. The problem of soliton transmission
and reflection is studied in \cite{Cascaval}through the numerical
solution of the stationary NLSE on graphs. 

Dispersion relations for linear and nonlinear NLSE on networks are discussed in \cite{Banica}.  More recent treatment of the stationary NLSE in the context of scattering from nonlinear
networks can be found in the Ref.\cite{Uzy4}. In particular, the
authors  discuss transmission through a
complex network of nonlinear one-dimensional leads and found the
existence of the high number of sharp resonances  dominating in the
scattering process. The stationary NLSE with power focusing
nonlinearity on star graphs was studied in very recent paper
\ci{Adami1}, where existence of the nonlinear stationary states
are shown for $\delta-$type boundary conditions.   
In particular, the authors of consider \ci{Adami1} a star graph with $N$ semi-infinite bonds, for which they obtain the exact solutions for the boundary conditions with $\alpha \neq 0$. The properties of the ground state wave function is also studied by considering separately the cases of odd and even $N$. 
We note that our work treats NLSE on simplest graphs with finite-length bonds,  aiming at obtaining its exact solutions for some types of the boundary conditions.

Despite the
fact that considerable interest to NLSE and soliton transport on
networks can be observed during last 2 years, the problem is still
very far from being studied comprehensively. In particular,
detailed treatment of the boundary conditions and exact solutions for
simplest topologies are not presented in yet in the literature. In this work we
present exact solutions of the stationary NLSE for three types of
simplest topologies, such as star, tree and loop graphs.

Motivation for the study of NLSE on graphs comes from the
different practically important applications such as soliton
transport in optical waveguide networks \cite{Burioni1},  soltion
dynamics in DNA double helix \cite{Yomosa}-\cite{Yakushevich} and
living systems \cite{Davydov} and other discrete structures
\cite{Burioni}.

An important applications of NLSE on networks is Bose-Einstein
condensation (BEC) and transport of BEC in networks. This issue that has been
extensively discussed recently in the literature
\cite{Leboeuf}-\cite{Oliv}. We note also that networks can be used a the traps for for BEC experiments.

It is important to note that earlier the problem of soliton transport in discrete structures and networks
was mainly studied within the discrete NLSE
\cite{Burioni}. However, such an approach doesn't provide
comprehensive treatment of the problem and one needs to use
continuous   NLSE on graphs. The aim of this work is the
formulation and solution of stationary NLSE on simplest graphs
such as star, tree and loop which can be considered as
exactly solvable topologies.

This paper is organized as follows. In the next section we will
present formulation of the problem for the primary star graph consisting of three bonds.
Section III presents derivation
of the exact solution for stationary NLSE on primary star graph by considering
repulsive and attractive nonlinearities. In section IV
we discuss the extension of these results to the case of other
simplest topologies such as tree graph, loop graph and their combinations.
Finally, section V presents some concluding remarks.

%%%%%%%%%%%%%%%%%%%%%%%%%%%%%%%%%%%%%%%%%%%%%%%%%%%%%%%%%%%%%%%%%%%

\section{Time-independent NLSE on star graphs}

The problem we want to treat is the stationary (time-independent)
NLSE with cubic nonlinearity  on the primary star graph. The star graph is a
three or more bonds connected to one vertex (branching point). The primary star graph consisting of three bonds, $b_1,\;b_2,\;b_3$,
is plotted in  Fig. 1.
The coordinate,  $x_1$ on the bond,
$b_1$ is defined from $0$ to $L_1$, while for the bonds  $b_k,
k=2,3$ the coordinates, $x_k$, are defined from $L_1$ to $L_k$. At
the branching point we have $x_k=L_1$. In the following we will
use notation $x$ instead of
 $x_k,\ (k=1,2,3)$. Then the time-independent NLSE
can be written for each bond as

\begin{figure}[htb]
\centerline{\includegraphics[width=70mm]{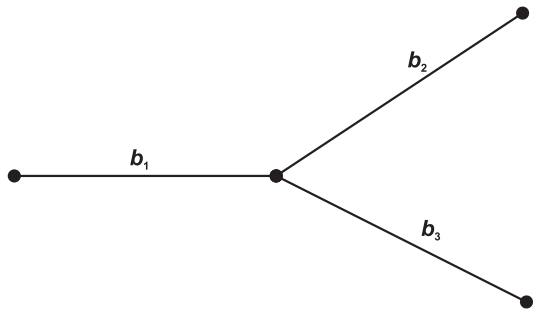}}
\caption{Primary star graph consisting of 3 bonds connected at a
vertex.} \label{simple-star}
\end{figure}

\begin{equation}
-\psi_j''+\beta_j|\psi_j|^2\psi_j=\lambda^2\psi_j,\;\beta_j>0,\;j=1,2,3.\label{eq4}
\end{equation}
Eq.\re{eq4} is a multi-component equation in which components are
mixed through the boundary conditions and conservation laws. Most simplest conservation
is the current conservation. For each bond the current is defined as
\begin{equation}
j_k(x)=2{\bf Im}\left[\psi_k^*(x)\frac{\partial\psi_k(x)}{\partial
x}\right].\nonumber
\end{equation}
This current should disappear at the ends of the each bond that can be
written in terms of the following conditions:
\begin{eqnarray}
\psi_1|_{x=0}=0,\;\frac{\partial\psi_1(x)}{\partial
x}\mid_{x=0}=\nu_1\psi_1(0),\nonumber\\
\psi_k|_{x=L_k}=0,\;\frac{\partial\psi_k(x)}{\partial
x}\mid_{x=L_k}=\nu_k\psi_k(L_k),\;k=2,3.\nonumber
\end{eqnarray}
The following local current conservation
condition (Kirchhoff law) should be valid at the branching point
\begin{equation}
j_1(L_1)=j_2(L_1)+j_3(L_1).\nonumber
\end{equation}

More detail analysis of the boundary conditions on graphs can be
found in the Refs. \ci{Exner1,Exner2}

In this paper we will consider the following boundary conditions:
\begin{eqnarray}
&\psi_1(L_1)=A_2\psi_2(L_1)=A_3\psi_3(L_1),&\nonumber\\
&\left[\frac{\partial}{\partial
x}\psi_1(x)-\frac{1}{A_2^*}\frac{\partial}{\partial
x}\psi_2(x)-\frac{1}{A_3^*}\frac{\partial}{\partial
x}\psi_3(x)\right]|_{x=L_1}=\alpha\psi_1(L_1),\ \ A_2A_3\not
=0.&\nonumber
\end{eqnarray}
In special case of  $A_2=A_3=1$, these boundary
conditions reproduce those considered in the Refs. \ci{Uzy1,Uzy2}.

%%%%%%%%%%%%%%%%%%%%%%%%%%%%%%%%%%%%%%%%%%%%%%%%%%%%%%%%%%%%%%%%%%%%%%%%%%%%%

\section{Solution of the time-independent NLSE on simplest graphs}

Consider the NLSE on the star  graph presented in Fig 1. with the
following boundary conditions ($\lambda$ is real;
$A_2=\sqrt{\beta_2/\beta_1},\; A_3=\sqrt{\beta_3/\beta_1}, \
\alpha=0$ )

\begin{eqnarray}
&\psi_1(x)|_{x=0}=0,\;\psi_{2,3}(x)|_{x=L_{2,3}}=0,&\label{eq8}\\
&\sqrt{\beta_1}\psi_1(L_1)=\sqrt{\beta_2}\psi_2(L_1)=\sqrt{\beta_3}\psi_3(L_1),&\label{eq9}\\
&\left[\frac{1}{\sqrt{\beta_1}}\frac{\partial}{\partial
x}\psi_1(x)-\frac{1}{\sqrt{\beta_2}}\frac{\partial}{\partial
x}\psi_2(x)-\frac{1}{\sqrt{\beta_3}}\frac{\partial}{\partial
x}\psi_3(x)\right]|_{x=L_1}=0,&\label{eq10}
\end{eqnarray}
where the wave function is normalized as follows:
\begin{equation}
\sum_{j=1}^3\int_{b_j}|\psi_j(x)|^2dx=1.\label{eq11}
\end{equation}

In the following we will consider Dirichlet boundary conditions at
the end-vertices of a graph,
 while at the branching points the boundary conditions given by Eqs.
(\ref{eq9}) and (\ref{eq10}) are to be considered. We note that in
the left-hand side of Eq. \ref{eq10} the derivatives are taken with
the plus sign, while in right-hand the signs of the derivatives are positive.
Furthermore, to make our notations different than those of the Refs. \ci{Adami, Adami1}, the
conditions given by Eqs. (\ref{eq9}), (\ref{eq10}) and their
generalizations will be called  $\delta''-$type conditions (the boundary conditions considered in the Refs.  \ci{Adami, Adami1} are denoted by  $\delta-$ and $\delta '$).

Solution of the NLSE can be written in the form
\begin{equation}
\psi_j(x)=exp(i\gamma_j(x))f_j(x),\;j=1,2,3,\label{eq12}
\end{equation}
where  $f_j(x)$ is a real function obeying the equation
\begin{equation}
-f_j''+\beta_jf_j^3=\lambda^2f_j.\label{eq13}
\end{equation}

Inserting  Eq. (\ref{eq12}) into  Eq. (\ref{eq4})we have
\begin{equation}
-i\gamma_j''f_j+\gamma_j'^2f_j'-f_j''+\beta_jf_j^3=\lambda^2f_j.\label{exeq1}
\end{equation}
Taking into account Eq. (\ref{eq13}) and requiring that real and
imaginary parts of Eq. should be zero (\ref{exeq1}) we get
$\gamma_j=const$.

The following relations can be obtained from Eq. (\ref{eq9})
\begin{equation}
exp(i\gamma_1)\sqrt{\beta_1}f_1(L_1)=exp(i\gamma_1)\sqrt{\beta_2}f_2(L_1)=exp(i\gamma_3)\sqrt{\beta_3}f_3(L_1).\nonumber
\end{equation}

The latter is valid only under the conditions
\begin{eqnarray}
&\gamma_1=\gamma_2=\gamma_3=\gamma,&\nonumber\\
&\sqrt{\beta_1}f_1(L_1)=\sqrt{\beta_2}f_2(L_1)=\sqrt{\beta_3}f_3(L_1).&\nonumber
\end{eqnarray}

It is clear that the functions  $f_1,f_2,f_3$ should obey Eq.
(\ref{eq8}) - (\ref{eq11}).

Exact solutions of Eq.(\ref{eq13}) for finite interval and periodic boundary conditions can be found in the Refs.  \cite{Car1,Car2}.
Here we consider this problem for the network boundary conditions given by Eq. (\ref{eq8}).
Partial solution
of Eq.~(\ref{eq13}) satisfying these boundary conditions can be
written as
$$
f_1(x)=B_1sn\left(\alpha_1x|k_1\right),
$$
$$
f_2(x)=B_2sn\left(\alpha_2(x-L_2)|k_2\right),
$$
$$
f_3(x)=B_3sn\left(\alpha_3(x-L_3)|k_3\right),
$$

where  $sn(ax|k)$ are the Jacobian elliptic functions.
\cite{Bowman}

The general solution can be written as
\begin{align}
f_b(x)=B_bsn\left(\alpha_bx+\delta_b|k_b\right).\nonumber
\end{align}

Inserting the last equation into Eq. (\ref{eq13}) and comparing
the coefficients of similar terms we have
\begin{eqnarray}
B_b=\sigma_b\sqrt{\frac{2}{\beta_b}}\alpha_bk_b,\ \ \
\lambda^2=\alpha_b^2\left(1+k_b^2\right),\label{eq21}
\end{eqnarray}
where $\sigma_b=\pm 1$.

Using the relations \cite{Bowman}
$$
\int_a^bsn^2\left(\alpha(x-c)|k\right)dx=\frac{1}{k^2}\int_a^b\left[1-dn^2\left(\alpha(x-c)|k\right)\right]dx=$$
$$
=\frac{1}{k^2}(b-a)-\frac{1}{\alpha
k^2}E\left[am\left(\alpha(b-c)\right)|k\right]+\frac{1}{\alpha
k^2}E\left[am\left(\alpha(a-c)\right)|k\right],
$$
$$
am(u+2K(k)|k)=\pi+am(u|k),
$$
$$
E(n\pi\pm\varphi|k)=2nE(k)\pm E(\varphi|k),
$$
and taking into account Eqs. (\ref{eq9})-(\ref{eq11}) and
(\ref{eq21}), we obtain the following nonlinear algebraic system
with respect to $\alpha_j$ and $k_j, \;j=1,2,3$:
\begin{align}\label{BC1}
\sqrt{\beta_1}B_1sn\left(\alpha_1L_1|k_1\right)=\sqrt{\beta_2}B_2sn\left(\alpha_2(L_1-L_2)|k_2\right)=\sqrt{\beta_3}B_3sn\left(\alpha_3(L_1-L_3)|k_3\right),
\end{align}

\begin{align}
\frac{B_1\alpha_1}{\sqrt{\beta_1}}cn\left(\alpha_1L_1|k_1\right)dn\left(\alpha_1L_1|k_1\right)-\frac{B_2\alpha_2}{\sqrt{\beta_2}}cn\left(\alpha_2(L_1-L_2)|k_2\right)dn\left(\alpha_2(L_1-L_2)|k_2\right)-\nonumber
\end{align}
\begin{align}\label{BC2}
-\frac{B_3\alpha_3}{\sqrt{\beta_3}}cn\left(\alpha_3(L_1-L_3)|k_3\right)dn\left(\alpha_3(L_1-L_3)|k_3\right)=0,
\end{align}
$$
\frac{B_1^2}{k_1^2}L_1+\frac{B_2^2}{k_2^2}(L_2-L_1)+\frac{B_3^2}{k_3^2}(L_3-L_1)=
$$
\begin{align}\label{NC1}
=1+\frac{B_1^2}{k_1^2\alpha_1}E\left[am(\alpha_1L_1|k_1)\right]+\frac{B_2^2}{k_2^2\alpha_2}E\left[am(\alpha_2(L_2-L_1)|k_2)\right]+\frac{B_3^2}{k_3^2\alpha_3}E\left[am(\alpha_2(L_3-L_1)|k_3)\right].
\end{align}

Here $K(k)$ and $E(k)$ are the complete elliptic integrals of
first and second kind, respectively.

In general case this system can be solved using the different  (e.g.,
Newton's or Krylov's method) iteration schemes.  However, below we will show
solvability of this system for two special cases.

 {\bf First case.}  Let
$\frac{2n_1+1}{L_1}=\frac{2n_{2,3}+1}{L_{2,3}-L_1}$, where
$n_{1,2,3}\in{\bf N}\cup\{0\}$.

Choosing
\begin{equation}
\alpha_1=-\frac{2n_1+1}{L_1}K(k_1),\;\alpha_{2,3}=\frac{2n_{2,3}+1}{L_{2,3}-L_1}K(k_{2,3}),\nonumber
\end{equation}
we have
\begin{equation}
-\alpha_1=\alpha_2=\alpha_3=\alpha,\; k_1=k_2=k_3=k, \ \
\sigma_1=\sigma_2=\sigma_3=1.\nonumber
\end{equation}

It is clear that under these conditions Eqs. (\ref{BC1}) and
(\ref{BC2}) are valid. Furthermore, it follows from Eq.(\ref{NC1})
that
\begin{equation}
g(k)\equiv
2\frac{(2n_1+1)^2}{L_1^2}\left(\frac{L_1}{\beta_1}+\frac{L_2-L_1}{\beta_2}+\frac{L_3-L_1}{\beta_3}\right)K(k)\left(E(k)-(1-k^2)K(k)\right)-1=0.\label{g1}
\end{equation}

The solvability of Eq.(\ref{g1}) is equavalent to that of NLSE on
graph. Therefore we will prove the solvability of this equation.
Since the following relations are valid
\begin{equation}
\lim_{k\to 0}g(k)=-1,\;\lim_{k\to 1}g(k)=+\infty,\nonumber
\end{equation}
 and $g(k)$  is a continuous function of $k$ on the interval
$(0;1)$ it follows that Eq.(\ref{g1}) has a root.

{\bf Second case.}  Now we show that there exist another solution
of Eq.(\ref{g1}). For the case when
$\alpha_1=\frac{(-1)^{n_1}p+2n_1K(k_1)}{L_1},\;\alpha_{2,3}=\frac{(-1)^{n_{2,3}}p+2n_{2,3}K(k_{2,3})}{L_{2,3}-L_1}$,
where $0<p<2K(k_{1,2,3})$, $n_{1,2,3}\in {\bf N}$ è $n_1,n_2,n_3$
cannot be odd or even at the same time,
from Eqs. (\ref{eq21}) and (\ref{BC1}) we can obtain
\begin{equation}
\alpha_1=\alpha_2=\alpha_3=\alpha,\; k_1=k_2=k_3=k, \ \sigma_1=1,\
\sigma_2=\sigma_3=-1.\nonumber
\end{equation}

From Eq. (\ref{BC2}) we have
\begin{equation}
\frac{(-1)^{n_1}}{\beta_1}+\frac{(-1)^{n_2}}{\beta_2}+\frac{(-1)^{n_3}}{\beta_3}=0.\nonumber
\end{equation}
Furthermore, it follows from the last equation and Eq.(\ref{NC1})
that
\begin{equation}
g(k)\equiv
4\left(\frac{(-1)^{n_1}p+2n_1K(k_1)}{L_1}\right)\left(\frac{n_1}{\beta_1}+\frac{n_2}{\beta_2}+\frac{n_3}{\beta_3}\right)\left(K(k)-E(k)\right)-1=0.\label{dreq3}
\end{equation}
We have
\begin{equation}
\lim_{k\to 0}g(k)=-1,\;\lim_{k\to 1}g(k)=+\infty.\nonumber
\end{equation}

Since  $g(k)$  is a continuous function of k on the interval
$(0;1)$, it follows from the last relations that Eq.(\ref{dreq3}) has a root.

%%%%%%%%%%%%%%%%%%%%%%%%%%%%%%%%%%%%%%%%%%%%%%%%%%%%%%%%%%%%%%%%%%%%%%%%%%%%%%%%%%%

\subsection{Case of attractive nonlinearity}

NLSE with attractive nonlinearity is of importance for a number of
problems of BEC physics \cite{Car2}. Earlier, it was treated for
the box-boundary conditions in the Ref. \cite{Car2}. Here we
extend the results of the Ref. \cite{Car2} to the case of the star
graphs. Consider also the case of NLSE with attractive
nonlinearity given as
\begin{equation}
-\psi_j''-\beta_j|\psi_j|^2\psi_j=\lambda^2\psi_j,\;\beta_j>0,\;j=1,2,3.\label{eq44}
\end{equation}

We write the solution of Eq. (\ref{eq44})  as

\begin{equation}
\psi_j(x)=B_jcn\left(\alpha_jx+\delta_j|k_j\right),\nonumber
\end{equation}
where the parameter $\delta_j$ can be determined  from the
boundary condition given by Eq. (\ref{eq8}) as
\begin{eqnarray}
\delta_1=(2m_1+1)K(k_1),\delta_j=(2m_j+1)K(k_j)-\alpha_jL_j,\; \ \
j=2,3,\nonumber
\end{eqnarray}
where $m_1,m_2,m_3\in {\bf Z}$. The coefficient  $B_j$ obeys  Eq.
(\ref{eq21}) with
\begin{equation}
\lambda^2=(1-2k_j^2)\alpha_j^2.\label{eq49}
\end{equation}

Taking into account Eqs. (\ref{BC1})-(\ref{NC1}) and (\ref{eq49})
we obtain the following nonlinear algebraic system with respect to
$\alpha_j,\;k_j\;j=1,2,3$:
\begin{align}
\sqrt{\beta_1}B_1cn\left(\alpha_1L_1+(2m_1+1)K(k_1)|k_1\right)=\sqrt{\beta_2}B_2cn\left(\alpha_2(L_1-L_2)+(2m_2+1)K(k_2)|k_2\right)=\nonumber\\
\sqrt{\beta_3}B_3cn\left(\alpha_3(L_1-L_3)+(2m_3+1)K(k_3)|k_3\right),\label{attreq1}\end{align}
\begin{align}
(1-2k_1^2)\alpha_1^2=(1-2k_2^2)\alpha_2^2=(1-2k_3^2)\alpha_3^2,\label{attreq2}\end{align}
\begin{align}
\frac{B_1\alpha_1}{\sqrt{\beta_1}}sn\left(\alpha_1L_1+(2m_1+1)K(k_1)|k_1\right)dn\left(\alpha_1L_1+(2m_1+1)K(k_1)|k_1\right)-\nonumber\\
\frac{B_2\alpha_2}{\sqrt{\beta_2}}sn\left(\alpha_2(L_1-L_2)+(2m_2+1)K(k_2)|k_2\right)dn\left(\alpha_2(L_1-L_2)+(2m_2+1)K(k_2)|k_2\right)-\nonumber\\
-\frac{B_3\alpha_3}{\sqrt{\beta_3}}sn\left(\alpha_3(L_1-L_3)+(2m_3+1)K(k_3)|k_3\right)dn\left(\alpha_3(L_1-L_3)+(2m_3+1)K(k_3)|k_3\right)=0,\label{attreq3}\end{align}
\begin{align}
\frac{B_1^2}{\alpha_1k_1^2}\left[E\left(am(\alpha_1L_1+(2m_1+1)K(k_1)|k_1)\right)-(2m_1+1)E(k_1)\right]+\nonumber\\
\frac{B_2^2}{\alpha_2k_2^2}\left[(2m_2+1)E(k_2)-E\left(am(\alpha_2(L_1-L_2)+(2m_2+1)K(k_2)|k_2)\right)\right]+\nonumber\\
\frac{B_3^2}{\alpha_3k_3^2}\left[(2m_3+1)E(k_3)-E\left(am(\alpha_3(L_1-L_3)+(2m_3+1)K(k_3)|k_3)\right)\right]=\nonumber\\
=1+B_1^2\frac{1-k_1^2}{k_1^2}L_1+B_2^2\frac{1-k_2^2}{k_2^2}(L_2-L_1)+B_3^2\frac{1-k_3^2}{k_3^2}(L_3-L_1).\label{attreq4}
\end{align}

Now we again consider two special cases .

{\bf First case.} Let
$\frac{2m_1+1}{L_1}=\frac{2m_{2,3}+1}{L_{2,3}-L_1}$, where
$m_{1,2,3}\in{\bf N}\cup\{0\}$.

Then choosing
\begin{equation}
\alpha_1=-\frac{2m_1+1}{L_1}K(k_1),\;\alpha_{2,3}=\frac{2m_{2,3}+1}{L_{2,3}-L_1}K(k_{2,3}).\label{attreq5},
\end{equation}
from Eqs. (\ref{attreq1}) and(\ref{attreq2}) we can obtain
\begin{equation}
-\alpha_1=\alpha_2=\alpha_3=\alpha,\; k_1=k_2=k_3=k,\ \
\sigma_1=\sigma_2=\sigma_3=1.\label{attreq6}
\end{equation}

It is clear that under these conditions Eq. (\ref{attreq3}) is
valid. Furthermore, it follows from Eq.(\ref{attreq6}) that
\begin{equation}
g(k)\equiv
2\frac{(2m_1+1)^2}{L_1^2}\left(\frac{L_1}{\beta_1}+\frac{L_2-L_1}{\beta_2}+\frac{L_3-L_1}{\beta_3}\right)K(k)\left(E(k)-(1-k^2)K(k)\right)-1=0.\label{attreq7}
\end{equation}

Again, the solvability of  Eq.(\ref{attreq7}) is equavalent to to
that of NLSE with attractive nonlinearity.

Therefore we will prove the existence of the roots of this
equation. We have
\begin{equation}
\lim_{k\to 0}g(k)=-1,\;\lim_{k\to 1}g(k)=+\infty.\nonumber
\end{equation}

Then the continuity of function $g(k)$ on the interval $(0;1)$
implies that Eq.(\ref{attreq7}) has a root.

{\bf Second case.} Now we show that there exist another solution
of Eq.(\ref{attreq7}). Let $\alpha_1=\frac{\pm
p+(4n_1-2m_1-1)K(k_1)}{L_1},\;\alpha_{2,3}=-\frac{\mp
p+(4n_{2,3}-2m_{2,3}-1)K(k_{2,3})}{L_{2,3}-L_1}$, where
$0<p<2K(k_{1,2,3})$, $n_{1,2,3}\in {\bf N}$ è
$4n_{1,2,3}>2m_{1,2,3}+1$.

Then we have
\begin{equation}
|\alpha_1|=|\alpha_2|=|\alpha_3|=\alpha,\; k_1=k_2=k_3=k, \ \
\sigma_1=1,\ \sigma_2=\sigma_3=-1.\label{attreq9}
\end{equation}

From the condition (\ref{attreq3}) we have
\begin{equation}
\frac{1}{\beta_1}-\frac{1}{\beta_2}-\frac{1}{\beta_3}=0.\nonumber
\end{equation}
Furthermore, from the last equation and normalization
condition we have the following relation:
\begin{equation}
g(k)\equiv 2\left|\frac{\pm
p+(4n_1-2m_1-1)K(k)}{L_1}\right|\left(\frac{4n_1-2m_1-1}{\beta_1}+\frac{4n_2-2m_2-1}{\beta_2}+\frac{4n_3-2m_3-1}{\beta_3}\right)\left(K(k)-E(k)\right)-1=0.\label{attreq10}
\end{equation}
Therefore, according to the asymptotical relations given by
\begin{equation}
\lim_{k\to 0}g(k)=-1,\;\lim_{k\to 1}g(k)=+\infty,\label{attreq11}
\end{equation}
and from the continuity of function   $g(k)$  in the interval $(0;1)$,
we can conclude that Eq.(\ref{attreq10}) has a root.

%%%%%%%%%%%%%%%%%%%%%%%%%%%%%%%%%%%%%%%%%%%%%%%%%%%%%%%%%%%%%%%%%%%%%%%%%%%%%5

\section{Other simplest graphs}

The above results can be extended for other types of
graphs, such as general star graphs,  tree  (Fig. 2),
loop (Fig. 3) and their combinations with appropriate
boundary and vertex conditions. This can be done using the same approach as that in the Ref. \cite{Zarif}.

\subsection{Tree Graph}
Here we will give details of such extension for the tree graph.

In the following  we assume that $\delta''-$
type boundary conditions are hold at the branching points, while at the end vertices
the wave function becomes zero.

\begin{figure}[htb]
\centerline{\includegraphics[width=80mm]{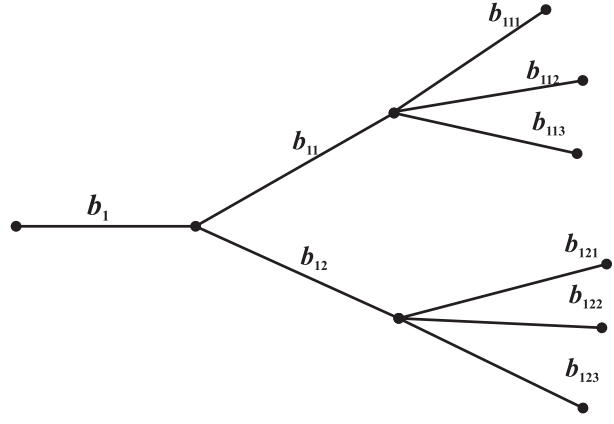}} \caption{Tree
 graph.} \label{tree}
\end{figure}

Furthermore, we seek for the solution of Eq. (\ref{eq13}) on the each bonds in
the form
\begin{align}
f_b=B_bsn\left(\alpha_bx+\delta_b\left|k_b\right.\right),\nonumber
\end{align}
where  $\delta_b$ are parameters that can be determined by the
given boundary conditions: $\delta_1=0,\
\delta_{1ij}=-\alpha_{1ij}L_{1ij}$.

Again, as it was done in the previous section for star graph,  one can show posibility of the exact solution
for  two special cases.

The first case is given by the relations
$$
\frac{4n_1+1}{L_1}=\frac{4(n_{1i}^{(2)}-n_{1i}^{(1)})}{L_{1i}-L_1}=\frac{4n_{1ij}+1}{L_{1ij}-L_{1i}},\;i=1,2,\;j=1,2,3
$$
where $n_1,n_{1i},n_{1ij}\in{\bf N}\cup\{0\},
n_{1i}^{(2)}>n_{1i}^{(1)}$ and
\begin{align}
\alpha_1=\frac{4n_1+1}{L_1}K(k_1),\;\alpha_{1i}=\frac{4(n_{1i}^{(2)}-n_{1i}^{(1)})}{L_{1i}-L_1}K(k_{1i}),\;\alpha_{1ij}=\frac{4n_{1ij}+1}{L_{1ij}-L_{1i}}K(k_{1ij}),
\delta_{1i}=\left(\frac{4(n_{1i}^{(1)}L_{1i}-n_{1i}^{(2)}L_{1})}{L_{1i}-L_{1}}+1\right)K(k_{1i}).\nonumber
\end{align}
It is easy to show that the following relations are valid:

\begin{equation}
\alpha_1=\alpha_{1i}=\alpha_{1ij}=\alpha,\; k_1=k_{1i}=k_{1ij}=k,\
\sigma_1=\sigma_{1i}=1,\ \sigma_{1ij}=-1.\nonumber
\end{equation}

Furthermore, we have
\begin{equation}
g(k)\equiv
2\frac{(4n_1+1)^2}{L_1^2}\left\{\frac{L_1}{\beta_1}+\sum_{i=1}^{2}\left[\frac{L_{1i}-L_1}{\beta_{1i}}+\sum_{j=1}^{3}\frac{L_{1ij}-L_{1i}}{\beta_{1ij}}\right]\right\}K(k)\left(K(k)-E(k)\right)-1=0.\nonumber
\end{equation}
Solvability of the last equation is obvious.

The second case is given by the relations
\begin{eqnarray}
&\alpha_1=\frac{-(-1)^{n_1}p_1+2n_1K(k_1)}{L_1},\;\alpha_{1i}=\frac{(-1)^{n_{1i}^{(2)}}p_{1i}-(-1)^{n_{1i}^{(1)}}p_{1}+2(n_{1i}^{(2)}-n_{1i}^{(1)})K(k_{1i})}{L_{1i}-L_1},\;\nonumber
\\
&\alpha_{1ij}=\frac{-(-1)^{n_{1ij}}p_{1i}+2n_{1ij}K(k_{1ij})}{L_{1ij}-L_{1i}},\nonumber\\
&\delta_{1i}=\frac{(-1)^{n_{1i}^{(1)}}p_{1}L_{1i}-(-1)^{n_{1i}^{(2)}}p_{1i}L_{1}+2(n_{1i}^{(1)}L_{1i}-n_{1i}^{(2)}L_{1})K(k_{1i})}{L_{1i}-L_1},\nonumber
\end{eqnarray}
where $0<p_1<2K(k_{1,1i}),\;0<p_{1i}<2K(k_{1i,1ij})$.

From vertex conditions we obtain
\begin{equation}
\alpha_1=\alpha_{1i}=\alpha_{1ij}=\alpha,\;
k_1=k_{1i}=k_{1ij}=k\nonumber
\end{equation}
and
\begin{equation}
\frac{(-1)^{n_{1}}}{\beta_{1}}-\sum_{i=1}^2\frac{(-1)^{n_{1i}^{(1)}}}{\beta_{1i}}=0,\;\frac{(-1)^{n_{1i}^{(2)}}}{\beta_{1}}+\sum_{j=1}^3\frac{(-1)^{n_{1ij}}}{\beta_{1ij}}=0.\nonumber
\end{equation}
It follows from the normalization condition that
\begin{equation}
g(k)\equiv
4\left(\frac{(-1)^{n_1}p+2n_1K(k)}{L_1}\right)\left(\frac{n_1}{\beta_1}+\sum_{i=1}^2\left[\frac{n_{1i}^{(2)}-n_{1i}^{(1)}}{\beta_{1i}}+\sum_{j=1}^3\frac{n_{1ij}}{\beta_{1ij}}\right]\right)\left(K(k)-E(k)\right)-1=0.\nonumber
\end{equation}
Solvability of the last equation follows from the properties of
the function $g(k)$.

\subsection{Graph With Loops}

\begin{figure}[htb]
\centerline{\includegraphics[width=80mm]{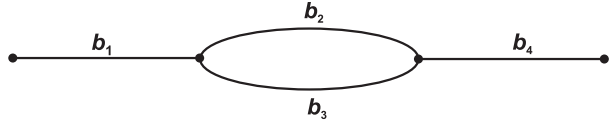}} \caption{Loop graph.} \label{loop}
\end{figure}

Similar results  can be obtained in for the loop graph plotted in
the Fig. \ref{loop}.

The solution of Eq.(\ref{eq13}) for this topology can be written
in terms of the Jacobian elliptic functions as
\begin{align}
f_1(x)=B_1sn\left(\alpha_1x|k_1\right),\nonumber\\
f_j(x)=B_jsn\left(\alpha_jx+\delta_j|k_j\right),\;j=2,3,\nonumber\\
f_4(x)=B_4sn\left(\alpha_4(x-L_4)|k_4\right).\nonumber
\end{align}

Again, two special cases will be considered.

For the first case we put
$\frac{4n_{1,4}+1}{L_{1,4}}=\frac{4(n_{2,3}^{(2)}-n_{2,3}^{(1)})}{L_{2,3}-L_1}$,
where $n_{1,4},n_{2,3}^{(1),(2)}\in{\bf N}\cup\{0\}$,
$n_{2,3}^{(2)}>n_{2,3}^{(1)}$. Choosing
\begin{eqnarray}
&\alpha_{1,4}=\frac{4n_{1,4}+1}{L_{}1,4}K(k_{1,4}),\;\alpha_{2,3}=\frac{4(n_{2,3}^{(2)}-n_{2,3}^{(1)})}{L_{2,3}-L_1}K(k_{2,3}),&\nonumber\\
&\delta_{2,3}=\left(\frac{4(n_{2,3}^{(1)}L_{2,3}-n_{2,3}^{(2)}L_{1})}{L_{2,3}-L_{1}}+1\right)K(k_{2,3}).&\nonumber
\end{eqnarray}

We have $\alpha_j=\alpha,\; k_j=k, j=1,2,3,4$,
$\sigma_1=\sigma_2=\sigma_3=1, \ \sigma_4=-1$. From the
normalization condition we get
\begin{equation}
g(k)\equiv
2\frac{(4n_1+1)^2}{L_1^2}\left\{\frac{L_1}{\beta_1}+\frac{L_{2}-L_1}{\beta_{2}}+\frac{L_{3}-L_1}{\beta_{3}}+\frac{L_{4}}{\beta_{4}}\right\}K(k)\left(K(k)-E(k)\right)-1=0.\nonumber
\end{equation}

The second case is given by
\begin{align}
&\alpha_1=\frac{(-1)^{n_1}p_1+2n_1K(k_1)}{L_1},\;\alpha_{2,3}=\frac{(-1)^{n_{2,3}^{(2)}}p_{2}-(-1)^{n_{2,3}^{(1)}}p_{1}+2(n_{2,3}^{(2)}-n_{2,3}^{(1)})K(k_{2,3})}{L_{2,3}-L_1},\nonumber\\
&\alpha_4=\frac{(-1)^{n_4}p_2+2n_4K(k_4)}{L_4},\nonumber\\
&\delta_{2,3}=\frac{(-1)^{n_{2,3}^{(1)}}p_{1}L_{2,3}-(-1)^{n_{2,3}^{(2)}}p_{2}L_{1}+2(n_{2,3}^{(1)}L_{2,3}-n_{2,3}^{(2)}L_{1})K(k_{2,3})}{L_{2,3}-L_1},\nonumber
\end{align}
where $0<p_1<2K(k_{1,2,3}),\;0<p_{2}<2K(k_{2,3,4})$.

Then from the $\delta''-$type vertex conditions we obtain
$\alpha_j=\alpha,\; k_j=k, j=1,2,3,4$ and
\begin{equation}
\frac{(-1)^{n_{1}}}{\beta_{1}}-\frac{(-1)^{n_{2}^{(1)}}}{\beta_{2}}-\frac{(-1)^{n_{3}^{(1)}}}{\beta_{3}}=0,\;\frac{(-1)^{n_{2}^{(2)}}}{\beta_{2}}+\frac{(-1)^{n_{3}^{(2)}}}{\beta_{3}}+\frac{(-1)^{n_{4}}}{\beta_{4}}=0.\nonumber
\end{equation}

Furthermore, according to the normalization condition one can obtain the following equation
\begin{equation}
g(k)\equiv
4\left(\frac{(-1)^{n_1}p+2n_1K(k)}{L_1}\right)\left(\frac{n_1}{\beta_1}+\sum_{i=2}^3\frac{n_{i}^{(2)}-n_{i}^{(1)}}{\beta_{i}}+\frac{n_{4}}{\beta_{4}}\right)\left(K(k)-E(k)\right)-1=0.\nonumber
\end{equation}

In both cases solvability of the equation $g(k)=0$ follows from
the asymptotical relations,  $\lim_{k\to 0}g(k)=-1,\;\lim_{k\to 1}g(k)=+\infty$
and from the continuity of the function $g(k)$.

\section{Conclusions}
In this paper we have treated the stationary nonlinear Schr\"odinger
equation with cubic nonlinearity for simplest graphs. Unlike the previous studies \cite{Zarif,Adami, Uzy4}, the lengths of the bonds are considered as finite. Therefore, our work can be considered as an extension of the results by L.D. Carr \ci{Car1,Car2, Car3} to the case of networks.
Formulation of the problem and its detailed treatment including method for
finding exact solutions are presented for the case
of star graph consisting of three bonds. The method is extended for
other simplest topologies such as tree and loop graphs. The cases
of attractive and repulsive nonlinearities are treated separately.
The above results can be useful for the problems of BEC on networks and discrete traps,
soliton transport optical waveguide networks,
soliton excitation in DNA double helix, energy transfer in nanoscale networks etc. In
principle, the method developed in this work can be extended for
more complicated topologies, too. Currently such a study is on progress.


\begin{thebibliography}{99}
\bibitem{Zakh1} Zakharov V E and Shabat Sov. Phys. JETP {\bf 34} 62 (1972)
\bibitem{Zakh2} Zakharov V E and Shabat  Funct. Anal. Appl. {\bf 8} 226 (1974)
\bibitem{Zakh3} Zakharov V E and Shabat  Funct. Anal. Appl. {\bf 13} 166 (1979)

\bibitem{Kivshar1} Y. S. Kivshar and G. P. Agarwal, {\it Optical Solitons: From Fibers
to Photonic Crystals} (Academic, San Diego, 2003).
\bibitem{Ablowitz} M.J. Ablowitz and P.A. Clarkson  {\it Solitons, Nonlinear Evolution Equations and Inverse Scattering}
(Cambridge: Cambridge University Press, 1999).
\bibitem{Pethick} C. J. Pethick and H. Smith, {\it Bose-Einstein Condensation in
Dilute Gases} (Cambridge University Press, Cambridge, England,
2002).
\bibitem{Lev} L.Pitaesvki and S. Stringari, {\it Bose-Einstein Condensation } (Oxford University Press, Oxford, England,
2003).
\bibitem{Thierry} Thierry Dauxois, Michel Peyrard, {\it Physics of Solitons} (Cambridge University Press, Cambridge, England,
2006).

\bibitem{Car1} L. D. Carr, Charles W. Clark, \& W. P.Reinhardt, Phys. Rev. A, {\bf 62}, 063610 (2000).
\bibitem{Car2} L. D. Carr, Charles W. Clark, \& W. P. Reinhardt, Phys. Rev. A,  {\bf 62}, 063611 (2000).
\bibitem{Well1} R. D'Agosta, B. A. Malomed, and C. Presilla, Phys. Lett. A {\bf 275}, 424, (2000).
\bibitem{Car3} L. D. Carr, K. W. Mahmud, and W. P. Reinhardt, Phys. Rev. A {\bf 64}, 033603 (2001).
\bibitem{Rap} K. Rapedius, D. Witthaut, and H. J. Korsch, Phys. Rev. A {\bf 73}, 033608 (2006).
\bibitem{Infeld} E.Infeld, P.Zin, J.Gocalek, \& M.Trippenbach, Phys. Rev. E,  {\bf 74}, 026610 (2006).

\bibitem{Harary} F. Harary, {\it Graph Theory} (Addison-Wesley, Reading, 1969).


\bibitem{Uzy1} Tsampikos Kottos and Uzy Smilansky, Ann.Phys., {\bf 76} 274 (1999).
\bibitem{Uzy2} Sven Gnutzmann and Uzy Smilansky, Adv.Phys. {\bf 55} 527 (2006).
\bibitem{Uzy3} S. GnutzmannJ.P. Keating b, F. Piotet, Ann.Phys., {\bf 325} 2595 (2010).
\bibitem{Exner1} P.Exner, P.Seba, P.Stovicek, J. Phys. A: Math. Gen. {\bf 21} 4009-4019 (1988).
\bibitem{Exner2} P.Exner, P.Seba,  Rep. Math. Phys., {\bf 28} 7 (1989).

\bibitem{Yomosa} S. Yomosa, Phys. Rev. A {\bf 27} , 2120 (1983).
\bibitem{Zhang} C.T.Zhang, Phys. Rev. A {\bf 35} , 886 (1987).
\bibitem{Yakushevich} L.V. Yakushevich, A.V. Savin, L.I. Manevitch, Phys. Rev. A {\bf 66} ,016614 (2002).

\bibitem{Davydov} A.S. Davydov, {\it Biology and Quantum Mechanics}, (Oxford: Pergamon, 1982)
\bibitem{Burioni} R. Burioni, D. Cassi, P. Sodano, A. Trombettoni, and A. Vezzani,
Chaos {\bf 15}, 043501 (2005); Physica D {\bf 216}, 71 (2006).

\bibitem{Leboeuf} P. Leboeuf and N. Pavloff, Phys. Rev. A {\bf 64} , 033602 (2001).
\bibitem{Bongs} K. Bongs, {\it et al}. Phys. Rev. A {\bf 63} , 031602 (R) (2001).
\bibitem{Stickney} J.A. Stickney, A.A. Zozulya, Phys. Rev. A {\bf 65}, 053612 (2002).
\bibitem{Stol}  O. Sotolongo-Costa1, G. J. Rodgers,  Phys. Rev. E {\bf 68} , 056118 (2003).
\bibitem{Paul1} T. Paul, P. Leboeuf, N. Pavloff,  K. Richter, and P. Schlagheck, Phys. Rev. A {\bf 72}, 063621 (2005).
\bibitem{Paul2} T. Paul, M. Hartung, K. Richter, and P. Schlagheck, Phys. Rev. A {\bf 76}, 063605 (2007).
\bibitem{Oliv} I. N. de Oliveira, Phys. Rev. E {\bf 81} , 030104(R) (2010).

\bibitem{Zarif} Z. Sobirov, D. Matrasulov, K. Sabirov, S. Sawada, and K. Nakamura, Phys. Rev. E {\bf 81} , 066602  (2010).
\bibitem{Adami} R.Adami, C.Cacciapuoti, D.Finco, D.N., Rev.Math.Phys, {\bf 23} 4 (2011).
\bibitem{Cascaval} R.C. Cascaval,  C.T. Hunter, Libertas Math. {\bf 30} 85 (2010).
\bibitem{Banica} V. Banica, L.Ignat, Arxiv: 1103.0429.
\bibitem{Uzy4} S. Gnutzmann, U. Smilansky, S. Derevyanko, Phys. Rev. A {\bf 83} 033831 (2011).
\bibitem{Adami1} R.Adami, C.Cacciapuoti, D.Finco, D.N., Arxiv: 1104.3839.

\bibitem{Burioni1} R. Burioni, D. Cassi, P. Sodano, A. Trombettoni, and A. Vezzani, Phys. Rev. E {\bf 73} , 066624  (2006).


\bibitem{Bowman} F. Bowman, {\it Introduction to Elliptic Functions, with Applications}
(Dover, New York, 1961).
\end{thebibliography}
\end{document}